\documentclass[12pt,preprint]{aastex}
\usepackage{amsmath}
\usepackage{graphicx}
\usepackage{ulem}
\begin{document}

\title{Constant cyclotron line energy in Hercules X--1 - \\Joint Insight-HXMT and NuSTAR observations}
\author{G. C. Xiao$^{1,2}$, L. Ji$^{3}$, R., Staubert$^{3}$, M. Y. Ge$^{1}$, S. Zhang$^{1,2}$, 
S. N. Zhang$^{1,2}$,  A. Santangelo$^{3}$, L. Ducci$^{3}$, J. Y. Liao$^{1}$, C. C. Guo$^{1,2}$
% ...
}

\affil{$^{1}$ Key Laboratory of Particle Astrophysics, Institute of High Energy Physics, Chinese Academy of Sciences, Beijing 100049, China.}
\affil{$^{2}$ University of Chinese Academy of Sciences, Chinese Academy of Sciences, Beijing 100049, China}
\affil{$^{3}$ Institut f\"ur Astronomie und Astrophysik, Sand 1, 72076 T\"ubingen, Germany}
\affil{$^{4}$ Department of Physics, Tsinghua University, Beijing 100084, China}
\affil{$^{5}$ Department of Astronomy, Beijing Normal University, Beijing 100088, China}
\affil{$^{6}$ Department of Engineering Physics, Tsinghua University, Beijing 100084, China}
\affil{$^{7}$ Key Laboratory of Space Astronomy and Technology, National Astronomical Observatories, Chinese
Academy of Sciences, Beijing 100012}

\begin{abstract}
The long-term evolution of the centroid energy of the CRSF in Her X-1 is still a mystery. 
We report a new measurement from a campaign between {\sl Insight}-HXMT and {\sl NuSTAR} 
performed in February 2018. Generally, the two satellites show well consistent 
results of timing and spectral properties. The joint spectral analysis 
confirms that the previously observed long decay phase has ended, and that 
the line energy instead keeps constant around 37.5 keV after flux correction.
\end{abstract}


\section{Introduction}
Her X-1 is a bright persistent X-ray binary (XRB) pulsar system, which 
was discovered in 1972 by Uhuru \citep{Tananbaum1972}. 
The system hosts a highly magnetized neutron star with a spin period 
of $\sim1.24$\,s, and a companion star with a mass 
of 2.2-{\it M}$_{\sun}$. The orbit is near circular with a high
inclination angle of $\sim$ 85\degr) causing a six hour long eclipse
within the 1.7-day orbital period. Her X-1 is the first source in which a 
cyclotron resonant scattering feature (CRSF) 
was detected \citep{Trumper1978}. The CRSF energy is around 40\,keV, 
varying with pulse phase, X-ray luminosity, with 35-d precession phase and with time\citep{Voges1982,Vasco2013,Staubert2007,Klochkov2011,Staubert2014}. 
Based on observations of {\sl RXTE}, {\sl Suzaku}, {\sl Integral} and {\it NuSTAR}, 
an almost 20 year long decay of the CRSF energy was observed 
since about 1996, ending somewhere between 2012 and 2015
\citep{Staubert2014,Staubert2016,Staubert2017,Klochkov2015,Ji2019}. 
Since then the CRSF energy appears to be constant 
around 37.5\,keV.

{\sl Insight}-HXMT, launched on June 15, 2017, was originally proposed in the 1990s, 
based on the Direct Demodulation Method \citep{Li93,Li94}. As the first X-ray astronomical 
satellite of China, {\it Insight-HXMT} carries three main instruments
\citep{Zhang2014}: the High Energy X-ray telescope (HE, 20-250\,keV, 5100\,cm$^{2}$), 
the Medium Energy X-ray telescope 
(ME, 5-30\,keV, 952\,cm$^{2}$), and the Low Energy X-ray telescope (LE, 1-15\,keV, 384\,cm$^{2}$). 
A series of observations for in-orbit calibrations were performed.
For instance, the instrument response was calibrated with observations 
of the Crab pulsar and the Crab nebular. Campaigns with other orbiting telescopes 
are part of the calibration strategy. One of such a campaign observing Her X-1 was carried out with  
{\sl NuSTAR} eight months after the launch of {\sl Insight}-HXMT. The aim of 
this campaign was to test the calibration of Insight-HXMT 
in both, the timing and the energy domains, with a focus on the energy 
channel relationship at energies above 10 keV by using the CRSF 
of Her X-1. In this work, we study the timing and spectral properties 
of Her X-1, and compare the results between {\sl Insight}-HXMT and 
{\sl NuSTAR}. 

\section{Observations and results}

\begin{figure}
\centering
\includegraphics[width=0.6\textwidth]{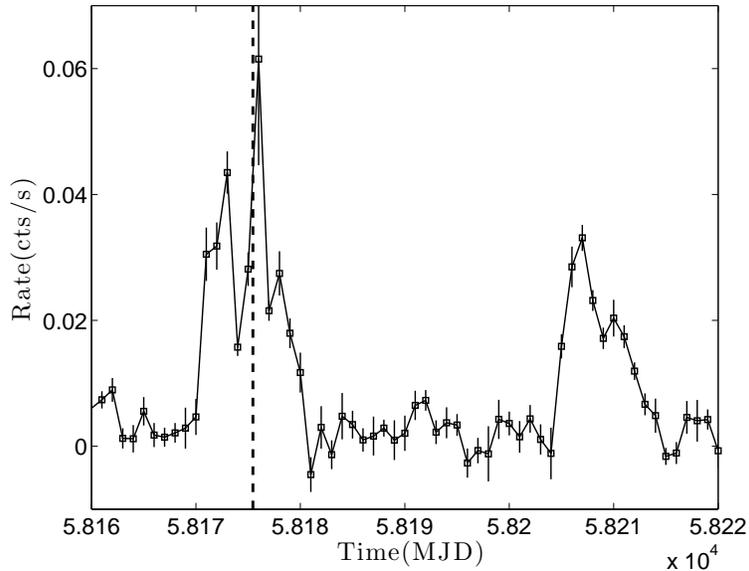}
\caption{The light-curve of Her X-1 as measured by \textit{Swift}-BAT. 
The vertical dashed line represents the time 
of the observations by {\sl Insight}-HXMT and {\sl NuSTAR}.
\label{fig1}}
\end{figure}

The campaign of Her X--1 observations with {\sl Insight}-HXMT 
and {\sl NuSTAR} was carried out on February 26th, 2018, around 
the peak flux of the Main-On of this 35\,d cycle  
($\sim$ 200\,mCrab detected with \textit{Swift}/BAT,
see Fig. \ref{fig1} and Table~\ref{table:exposure}).

We performed the data reduction by using the {\sl Insight}-HXMT data 
analysis software package {\sc HXMTDas}(v2.01) and followed
the recommended procedures \footnote{For details, 
see http://www.hxmt.org/index.php/enhome/analysis/199-hxmt-data-anslysis-software}.
We calibrated the raw events by using routines \texttt{hepical}, 
\texttt{mepical} and \texttt{lepical} for HE, ME and LE instruments, and 
estimated the corresponding Good Time Intervals (GTIs), based on the following 
criteria: the earth elevation angle $>$15$^{\circ}$ 
(10$^{\circ}$) for LE (HE and ME); the cutoff rigidity (COR) $>$ 8$^{\circ}$; 
the offset angle is smaller than 0.04; discarding the data near 
the South Atlantic Anomaly (SAA) passage; the Bright Earth Angle is $>$ 40$^{\circ}$ for LE.
In our analysis of Her X-1 the background model of {\sl Insight}-HXMT is 
generated with a large amount of blank sky observations \citep{bkg}.
The background was calculated by using standalone {\sc python} 
scripts \texttt{hebkgmap}, \texttt{mebkgmap} and \texttt{lebkgmap} 
\citep[for details, see ][]{LiCal2018}.
The response files for the spectral analysis were constructed according to the CALDB version {2.01}.

\textit{NuSTAR} is the first space-based directly imaging X-ray telescope at 
energies above 10\,keV \citep{Harrison2013}. It consists of two 
focal plane module telescopes (FPMA and FPMB). We utilised the standard 
software \texttt{NUSTARDAS} included in \textsc{HEASoft} 
(version 6.24) to perform the data reduction. During the spectral analysis, we 
extracted the source spectrum within a circular region, i.e.,  
$180^{\prime\prime}$ centred on the pulsar, and the background spectrum 
within a radius of 150$^{\prime\prime}$ which does not enclose 
(but close to) the source.

In the timing analysis, we made the barycentric correction with official tools, i.e., 
\texttt{hxbary} and \texttt{barycorr} developed for 
\textit{Insight}-HXMT and \textit{NuSTAR}, respectively. The binary correction was performed
by using the orbital ephemeris reported by \citep{Staubert2009}.
The spin period and its derivatives were obtained by using the phase-connection technique \citep{Deeter1981}.
In practice, we divided the observations into several segments (200\,s each), and 
searched for corresponding spin frequencies using the 
epoch folding method. We then folded the events to obtain pulse profiles for each 
segment, and derived an averaged pulse profile by 
co-aligning and combining all. The time-of-arrivals (TOAs) of pulsations were 
calculated by cross-correlating the pulse profile in each 
segment with the averaged pulse profile. The errors of TOAs were 
estimated by using Monte-Carlo method \citep{Ge2012}.
Finally, according to TOAs, we fitted the frequency and its derivative 
of Her X-1 jointly using the software {\sc TEMPO2} \citep{TEMPO2}.

\begin{figure}
\centering
\includegraphics[width=0.6\textwidth]{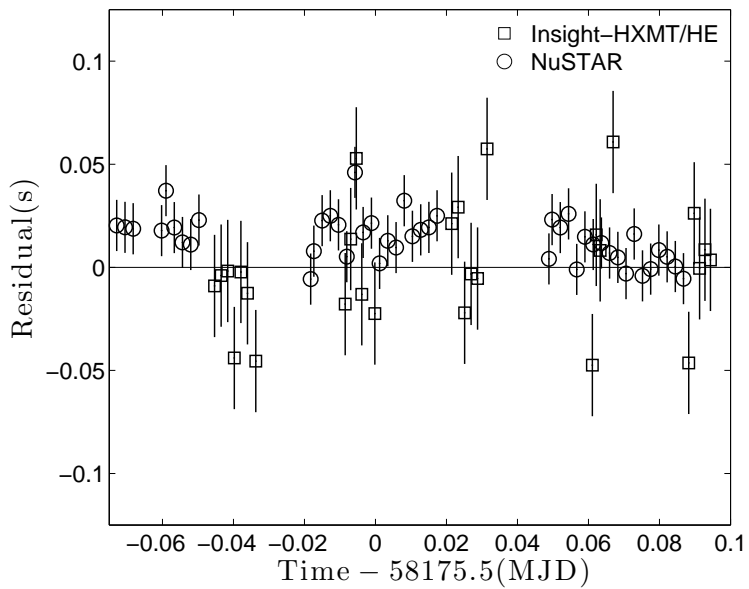}
\caption{The timing residuals in seconds of {\sl Insight}-HXMT/HE and {\sl NuSTAR} by using the parameters listed in 
Table \ref{table:para}. The square and circle points represent the data from {\sl Insight}-HXMT/HE and {\sl NuSTAR}, respectively. 
\label{fig2}}
\end{figure}

\begin{figure}
\centering
\includegraphics[width=0.6\textwidth]{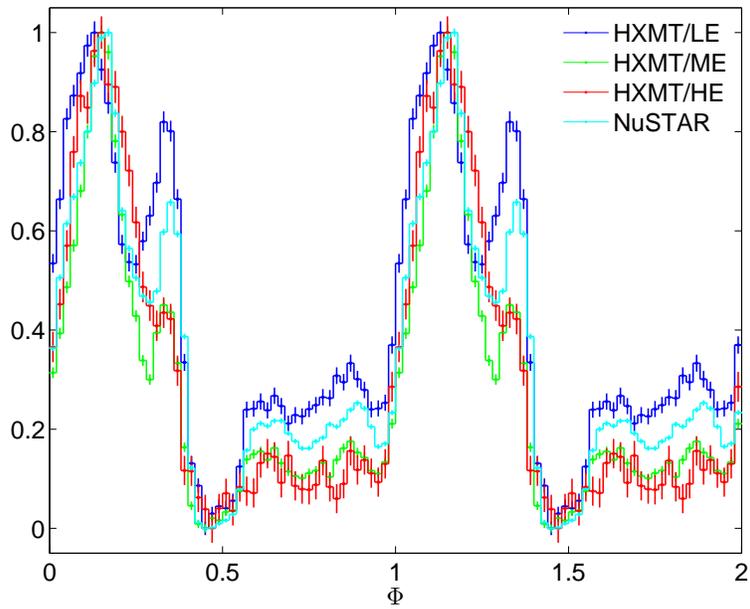}
\caption{The observed pulse profiles. 
Blue, green and red lines show pulse profiles observed with LE (1-10\,keV), 
ME (10-30\,keV) and HE(30-70\,keV) of {\sl Insight-HXMT}, respectively. 
The cyan line represents the pulse profile observed with {\sl NuSTAR} (3-70\,keV). 
\label{fig3}}
\end{figure}

We show the timing residuals after fitting \textit{Insight}-HXMT/HE's and NuSTAR's TOAs in Fig.~\ref{fig2}.
The resulting spin frequency is shown in Table~\ref{table:para}. The timing  system of {\sl Insight}-HXMT is well within the precision level of 15\,ms, which is consistent  with those of the timing residuals 
derived from timing analysis of the Crab pulsar \citep{LiCal2018}.
We folded the events observed with different instruments to obtain pulse profiles at different energies (Figure~\ref{fig3}). 
The pulse profiles consist of a main pulse with several smaller structures, which are similar to those of previous \textit{NuSTAR} 
observations \citep{Furst2013}.  In addition, pulse profiles are clearly energy-dependent, which will be extensively investigated in a 
separated paper. 

In the spectral analysis, we only considered the observational intervals outside of the eclipse.
During fittings, we employed a canonical "highecut" model, i.e., a power 
law model with a high energy cutoff, which is widely used for 
Her X-1 \citep[e.g.,][]{Staubert2017}. To smooth the phenomenological 
model around the cutoff energy, we multiplied the model with a 
Gaussian optical-depth profile ("gabs" in {\sc XSPEC}) with its energy tied to the cutoff energy \citep{Furst2013}.
A Gaussian model was added for the iron $K_{\rm \alpha}$ emission line around 6.4\,keV.
To investigate the absorption feature of the cyclotron line, another "gabs" component was considered.
In the spectral analysis, we used {\it XSPEC} (Version 12.10.0c) \citep{arnaud1996xspec}.
A systematic error of {2\%} was included for spectral fitting of  Insight-HXMT.
All uncertainties quoted in this paper are given at the $1\sigma$ (68.3\%) confidence level.

\begin{figure}
\centering
\includegraphics[width=0.8\textwidth]{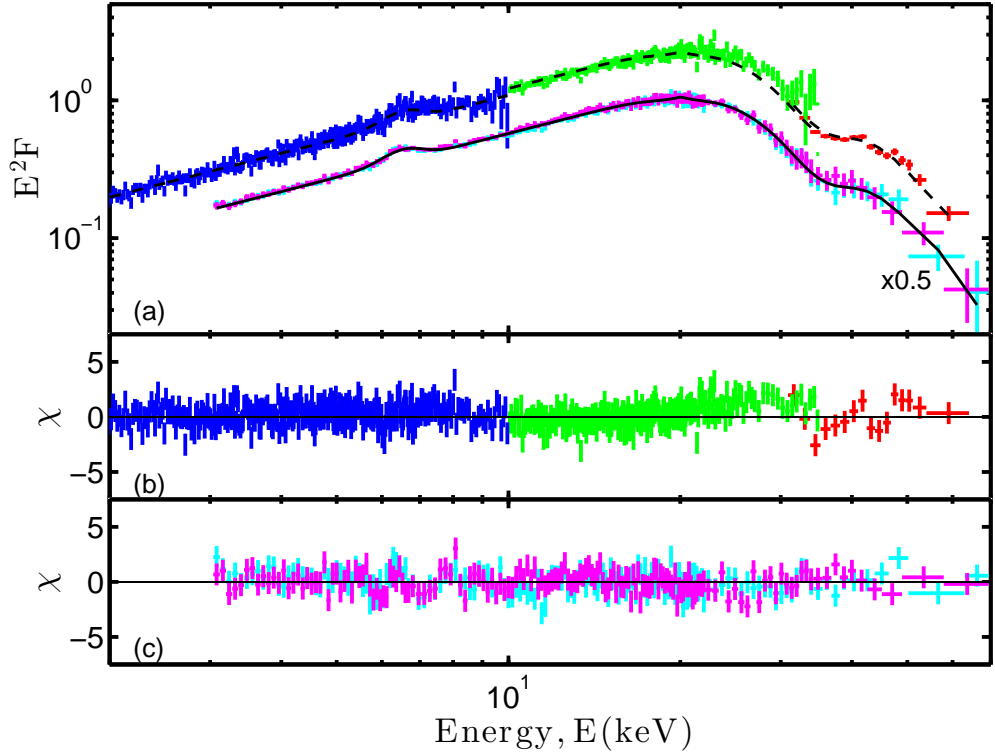}
\caption{The observed spectra of Her X-1. Panel (a): The black dashed and thin lines represent the spectral model. The blue, green 
and red points represent the observations of LE, ME and HE \textbf{of {\sl Insight-HXMT}}, respectively. The cyan and purple points are 
results of {\sl NuSTAR}/FPMA and FPMB, respectively. For clarity, the {\sl NuSTAR} observations are multiplied by a factor of 0.5.
%in order to show them separately.
Panel (b): The spectral residuals of LE, ME and HE \textbf{of {\sl Insight-HXMT}}, respectively. 
Panel (c): The spectral residuals of {\sl NuSTAR}/FPMA and FPMB, respectively.
\label{fig4}}
\end{figure}

We firstly fitted the spectra observed by \textit{Insight}-HXMT and \textit{NuSTAR} 
independently, using the model described above. We show the best-fitting 
parameters in Table~\ref{table:spec}. Clearly, the parameters obtained from the two satellites are generally well 
in agreement with each other, which verifies the performance of the in-orbital calibration of \textit{Insight}-HXMT.
However, the resulting "norm" parameter shows a $\sim$ 10\% discrepancy, which might be due 
to the influence of soft X-rays ($\le$ 3\,keV) of \textit{Insight-HXMT} that could influence the spectral shape significantly.
Thus, for comparison, we ignored the energy range below 3\,keV for \textit{Insight-HXMT} 
data, and found that in this case this discrepancy 
disappeared. We then jointly fitted \textit{Insight}-HXMT and \textit{NuSTAR} spectra, 
and obtained an acceptable  goodness-of-fit (shown in Figure~\ref{fig4}), demonstrating 
again the consistency of the cross-calibration between \textit{Insight}-HXMT and \textit{NuSTAR}.
We find that the spectral parameters in our observations are quite similar to 
previous reports \citep[see, e.g.,][]{Furst2013}. We also performed a spectral 
analysis of the {\sl NuSTAR} observation during the eclipse. We find that the CRSF centroid energy is 
derived consistent with that from the pre-eclipse {\sl NuSTAR} data. 

\begin{figure}
\centering
\includegraphics[width=0.8\textwidth]{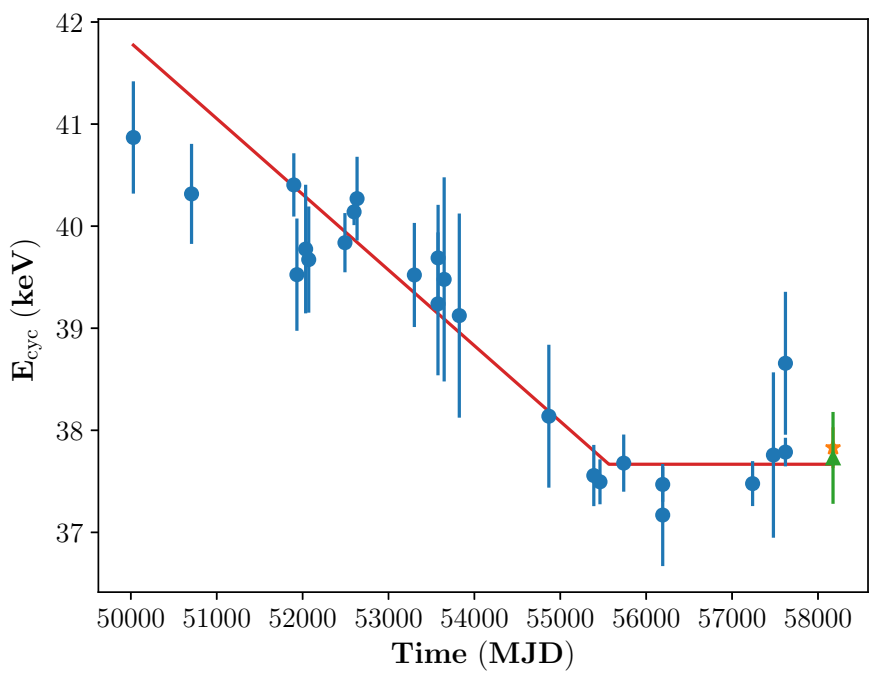}
\caption{Evolution of the flux corrected CRSF centroid energy of Her X-1. The blue points are
original measurements reported in \citet{Staubert2007,Staubert2014, Staubert2016, Staubert2017},
flux corrected by 0.68\,keV per ASM-cts/s. The green and yellow
points are the results 
of this work from {\sl Insight}-HXMT and {\sl NuSTAR}, respectively.
The evolution is described as a piecewise function, i.e., the red line. 
(See also \citet{Ji2019}).
\label{fig5}}
\end{figure}

We update the long-term evolution of the CRSF energy 
in Figure~\ref{fig5}. We use the originally measured values 
from \citet{Staubert2014, Staubert2016, Staubert2017}, corrected
for the dependence on flux using a correction factor of 0.68\,keV 
per ASM-cts/sec (this is the best currently available value,
determined by Staubert) \citep[for details, see, e.g.,][]{Staubert2016, Staubert2017,Ji2019}.
Clearly, the decay trend is stopped and afterwards the line energy stays constant.
We used a piecewise function to describe the evolution, and found that the averaged 
flux corrected line energy is 37.58$\pm$0.07\,keV for the time after MJD 55400.
We note that a F-test suggests that this function is significantly better than a simple 
power-law function with a p-value of $2.3\times10^{-7}$, and therefore the ending 
of the decreasing trend of the CRSF energy at a significance level of $>$5\,$\sigma$ 
for the first time. This result, including our new observations, is well consistent with previous 
reports \citep{Staubert2016,Staubert2017,Ji2019}.

\section{Summary}\label{sec:con}

We have performed the joint timing and spectral analyses of data on Her X-1 by using the contemporary observations by
{\sl Insight}-HXMT and {\sl NuSTAR} of February 2018. The results show that the pulse profiles of {\sl Insight}-HXMT 
are well consistent with those from {\sl NuSTAR}. The spectral parameters as derived from {\sl Insight}-HXMT are also 
consistent with those from {\sl NuSTAR}. We find that the centroid energy of the CRSF keeps the same trend as reported in the 
literature \citep{Staubert2016,Staubert2017,Ji2019}, i.e., being stable at $\sim$ 37.6\,keV after the long-term decay phase.

\begin{acknowledgements}
This work is supported by the National Key R\&D Program of China (2016YFA0400800) 
and the National Natural Science Foundation of China under grants 11673023, U1838201, U1838202 and U1838104. 
This work made use of the data from the HXMT mission, a project funded by China National Space Administration (CNSA) 
and the Chinese Academy of Sciences (CAS) and from {\sl NuSTAR}, an X-ray satellite operated by NASA.
\end{acknowledgements}

\clearpage

\clearpage

\begin{table}
\caption{Exposure times of the pointed observations of Her X--1}
\label{table:exposure}
%\medskip
\centering
\begin{tabular}{lcllr}
\hline\hline
Instrument  & Obs ID &      Date    &  Exposure (ks) \\ 
                   &            &                   &  ks \\ 
\hline
{\sl Insight}-HXMT  & P010130801006 & 2018/2/26  & 10 \\
{\sl NuSTAR}     & 30302012004 & 2018/2/26 & 62 \\
\hline
\end{tabular}
\end{table}

\begin{table}[htbp]
\caption{Pulse frequency of Her X--1}
\scriptsize{} \label{table:para}
\begin{center}
\begin{tabular}{ llllllc}
\hline\hline
&Time range (MJD)                         & 58175.45--58175.6    \\
& PEPOCH(MJD)                            & 58175.45  \\
&$\nu$(Hz)                                      & 0.807938(2)           \\
& rms(\,ms )                                     & 15            \\
& $\chi^2$/d.o.f. (d.o.f)                     & 1.01 (60) \\
\hline
\end{tabular}
\end{center}
\end{table}

\begin{table}[htbp]
\caption{Spectral parameters of Her X--1}
\scriptsize{} \label{table:spec}
\begin{center}
\begin{tabular}{ llllllc}
\hline\hline
&Parameters                              &      {\sl NuSTAR}         &  {\sl Insight}-HXMT(*)   &  Both                      &  {\sl Insight}-HXMT(**)  \\
& ${\rm E_{CRSF}}$\,(keV)        & $35.8\pm0.2$             &   $35.7\pm0.5$             &  $35.8\pm0.2$       &  $35.7^{+0.5}_{-0.4}$  \\
& $\sigma_{\rm CRSF}$\,(keV)  & $4.8\pm0.2$               &   $3.7\pm0.8$               &  $4.7\pm0.2$         &  $3.6^{+1.}_{-0.8}$ \\
& Strength CRSF                       & $6.0\pm0.4$               &   $6\pm1$                      & $6.2\pm0.4$          &  $5.5^{+1.9}_{-1.1}$ \\
& ${\rm E_{smooth}}$\,(keV)      & $19.8$                        &   $21.2$                         &  $19.9$                 &  $21.4$  \\
& $\sigma_{\rm smooth}$\,(keV) & $2.0\pm0.1$              &  $2.9^{+0.6}_{-0.5}$      &  $2.0\pm0.1$         & $2.5\pm0.6$\\
& Strength smooth                      & $0.62\pm0.05$           &  $1.1^{+0.4}_{-0.3}$     & $0.64\pm0.05$      & $0.8^{+0.5}_{-0.3}$\\
& ${\rm E_{cut}}$\,(keV)              & $19.8\pm0.1$             &  $21.2^{+0.9}_{-0.6}$   & $19.9\pm0.1$        & $21.4^{+1.4}_{-0.6}$ \\
& ${\rm E_{fold}}$\,(keV)             &  $9.6\pm0.1$              &  $9.6\pm0.4$                & $9.65\pm0.08$      & $9.8\pm0.4$         \\
&  $\Gamma$                              & $0.955\pm0.003$       &  $0.89\pm0.01$            & $0.951\pm0.002$  & $0.94\pm0.03$ \\
& Norm($10^{-3}$)                      & $102.0\pm0.6$            &  $88\pm1$                    & $96.3\pm0.5$        & $97\pm4$   \\
& ${\rm E_{Fe}}$\,(keV)               & $6.46\pm0.01$           &  $6.61\pm0.08$            & $6.45\pm0.01$      & $6.62\pm0.08$\\
& ${\rm \sigma_{Fe}}$\,(keV)       & $0.48\pm0.02$           &  $0.5\pm0.1$                & $0.52\pm0.02$      &  $0.7\pm0.2$ \\
& ${\rm Norm_{Fe}}$\,($10^{-3}$) & $4.1\pm0.1$              &  $4.1^{+0.9}_{-0.8}$     & $4.2\pm0.1$          &  $5.5^{+1.6}_{-1.3}$\\
& $\chi^2/{\rm d.o.f} (d.o.f)$          & 1.14(1671)                 &  0.76(1468)                   & 1.03(3155)            &  0.75(1261)  \\
%& $\chi^2({\rm d.o.f})$              & 1.14(1671)                    &    0.76(1468)             & 1.03(3155)            &    0.75(1261)  \\
\hline
\end{tabular}
\end{center}
* represents the spectrum fitting from 1.2\,keV.
** represents the spectrum fitting from 3\,keV.
\end{table}

\end{document}